\documentclass[conference, 9pt]{IEEEtran}
\usepackage{caption}
\usepackage{subcaption}
\usepackage{nopageno}
\usepackage{graphbox} 
\usepackage{amsmath,amsfonts,amsbsy,latexsym}
\usepackage{multirow}
\usepackage{makecell}
\usepackage{xcolor}
\usepackage{colortbl}
\usepackage{graphicx}
\usepackage{tabulary}
\usepackage{url}
\usepackage{booktabs}
\usepackage{cite}

\usepackage{smartdiagram}

\usepackage[activate={true,nocompatibility},
            final,
            tracking=true,
            kerning=true,
            spacing=true,
            factor=1100,
            stretch=10,
            shrink=10]{} 

\usepackage[english]{babel}
\usepackage{blindtext}
\usepackage{helvet}
\usepackage[T1]{fontenc}

\usepackage{booktabs}
\usepackage{tabu}

\usepackage{amsmath}
\usepackage{amssymb}
\usepackage{amsthm}
\usepackage{algorithmic}
\usepackage{algorithm}
\usepackage{siunitx}
\usepackage{amsmath,amsfonts,amsthm,mathrsfs, mathtools}
\usepackage{multirow}
\usepackage{parallel}

\usepackage{marvosym}
\usepackage{graphicx}
\usepackage{pgfpages}
\graphicspath{{images/}{../images/}}
\usepackage{pgffor}
\usepackage{xcolor}
\usepackage{colortbl}
\usepackage{pgfplots}
\usepackage{tikz}
\usepackage{tikzsymbols}
\usetikzlibrary{automata,positioning,fadings,shadows, shapes.arrows, shapes.geometric}
\usetikzlibrary{matrix, chains, patterns}
\usetikzlibrary{arrows,shapes.gates.logic.US,shapes.gates.logic.IEC,calc,automata}
\usepgfplotslibrary{groupplots}

\usepackage{pifont}

\definecolor{Paired-2}{RGB}{166,206,227}
\definecolor{Paired-1}{RGB}{31,120,180}
\definecolor{Paired-4}{RGB}{178,223,138}
\definecolor{Paired-3}{RGB}{51,160,44}
\definecolor{Paired-6}{RGB}{251,154,153}
\definecolor{Paired-5}{RGB}{227,26,28}
\definecolor{Paired-8}{RGB}{253,191,111}
\definecolor{Paired-7}{RGB}{255,127,0}
\definecolor{Paired-10}{RGB}{202,178,214}
\definecolor{Paired-9}{RGB}{106,61,154}
\definecolor{Paired-12}{RGB}{255,255,153}
\definecolor{Paired-11}{RGB}{177,89,40}

\usepackage{tikz-timing}[2009/05/15]

\usepackage{scalerel} 

    \pgfdeclarelayer{background}
    \pgfdeclarelayer{foreground}
    \pgfsetlayers{background,main,foreground}
    \tikzstyle{vblock} = [draw,
                          fill=dateorange!30,
                          rotate=90,
                          anchor=north west,
                          minimum width=3.5cm,
                          minimum height=.5cm,
                         rounded corners=.05cm,
                          ]

    \tikzstyle{tile} = [draw,
                        fill=dateblue!30,
                        anchor=south west,
                        minimum width=3.5cm,
                        minimum height=.5cm,
                        rounded corners=.05cm,
                        ]

    \tikzstyle{nnlayer} = [draw,
                        fill=dateblue!30,
                        anchor=south west,
                        minimum width=1cm,
                        minimum height=.5cm,
                        rounded corners=.05cm,
                        ]

    \tikzstyle{petile} = [draw,
                        fill=dateorange!30,
                        anchor=south west,
                        minimum width=1.5cm,
                        minimum height=.5cm,
                        rounded corners=.05cm,
                        ]

    \tikzstyle{arrow} = [->,
                         >=stealth,
                         thick,
                         rounded corners=.1cm,
                         color=datemagenta,
                         ]

    \tikzstyle{arrowg} = [->,
                          >=stealth,
                          thick,
                          rounded corners=.1cm,
                          color=lightgray,
                          ]
    \tikzset{%
    table/.style={%
      matrix of nodes,
      row sep=-\pgflinewidth,
      column sep=-\pgflinewidth,
      nodes={rectangle,draw=black,text width=1.25ex,align=center},
      text depth=0.25ex,
      text height=1ex,
      nodes in empty cells
      },
    texto/.style={font=\footnotesize\sffamily},
    title/.style={font=\small\sffamily}
    }

    \pgfmathdeclarefunction{gauss}{2}{%
        \pgfmathparse{1/(#2*sqrt(2*pi))*exp(-((x-#1)^2)/(2*#2^2))}
    }

    \pgfmathdeclarefunction{gaussprune}{2}{%
        \pgfmathparse{or(x<-1,x>1)/(#2*sqrt(2*pi))*exp(-((x-#1)^2)/(2*#2^2))}
    }

\usepackage{stfloats}

\usepackage{perpage}
\usepackage{hyperref}
\usepackage{csquotes}

\usepackage{pgfpages}

\makeatother

\definecolor{dateblue}{RGB}{36,80,117}
\definecolor{datemagenta}{RGB}{183,50,101}
\definecolor{dateorange}{RGB}{198,84,54}
\definecolor{datebrown}{RGB}{198,140,54}
\definecolor{dateyellow}{RGB}{198,178,54}

\tikzfading[name=arrowfading, top color=transparent!0, bottom color=transparent!95]
\tikzset{arrowfill/.style={#1, general shadow={fill=black, shadow yshift=-0.8ex, path fading=arrowfading}}}
\tikzset{arrowstyle/.style n args={3}{draw=#2,arrowfill={#3}, single arrow,minimum height=#1, single arrow,
single arrow head extend=.3cm,}}

\NewDocumentCommand{\tikzfancyarrow}{O{2cm} O{dateblue} O{top color=dateblue!20, bottom color=dateblue} m}{
\tikz[baseline=-0.5ex]\node [arrowstyle={#1}{#2}{#3}] {#4};
}

\usepackage{circuitikz}

{
\begin{tikzpicture}[baseline=(current bounding box.north),scale=0.8]
\draw (0,0) grid (4,4);
\draw (0,4) -- node [pos=0.7,above right,anchor=south west] {$x_3x_4$} node [pos=0.7,below left,anchor=north east] {$x_1x_2$} ++(135:1);
\matrix (mapa) [matrix of nodes,
        column sep={0.8cm,between origins},
        row sep={0.8cm,between origins},
        every node/.style={minimum size=0.3mm},
        anchor=8.center,
        ampersand replacement=\&] at (0.5,0.5)
{
                       \& |(c00)| 00         \& |(c01)| 01         \& |(c11)| 11         \& |(c10)| 10         \& |(cf)| \phantom{00} \\
|(r00)| 00             \& |(0)|  \phantom{0} \& |(1)|  \phantom{0} \& |(3)|  \phantom{0} \& |(2)|  \phantom{0} \&                     \\
|(r01)| 01             \& |(4)|  \phantom{0} \& |(5)|  \phantom{0} \& |(7)|  \phantom{0} \& |(6)|  \phantom{0} \&                     \\
|(r11)| 11             \& |(12)| \phantom{0} \& |(13)| \phantom{0} \& |(15)| \phantom{0} \& |(14)| \phantom{0} \&                     \\
|(r10)| 10             \& |(8)|  \phantom{0} \& |(9)|  \phantom{0} \& |(11)| \phantom{0} \& |(10)| \phantom{0} \&                     \\
|(rf) | \phantom{00}   \&                    \&                    \&                    \&                    \&                     \\
};
}%
{
\end{tikzpicture}
}

{
\begin{tikzpicture}[baseline=(current bounding box.north),scale=0.8]
\draw (0,0) grid (4,2);
\draw (0,2) -- node [pos=0.7,above right,anchor=south west] {$x_2x_3$} node [pos=0.7,below left,anchor=north east] {$x_1$} ++(135:1);
\matrix (mapa) [matrix of nodes,
        column sep={0.8cm,between origins},
        row sep={0.8cm,between origins},
        every node/.style={minimum size=0.3mm},
        anchor=4.center,
        ampersand replacement=\&] at (0.5,0.5)
{
                      \& |(c00)| 00         \& |(c01)| 01         \& |(c11)| 11         \& |(c10)| 10         \& |(cf)| \phantom{00} \\
|(r00)| 0             \& |(0)|  \phantom{0} \& |(1)|  \phantom{0} \& |(3)|  \phantom{0} \& |(2)|  \phantom{0} \&                     \\
|(r01)| 1             \& |(4)|  \phantom{0} \& |(5)|  \phantom{0} \& |(7)|  \phantom{0} \& |(6)|  \phantom{0} \&                     \\
|(rf) | \phantom{00}  \&                    \&                    \&                    \&                    \&                     \\
};
}%
{
\end{tikzpicture}
}

{
\begin{tikzpicture}[baseline=(current bounding box.north),scale=0.8]
\draw (0,0) grid (4,2);
\draw (0,2) -- node [pos=0.7,above right,anchor=south west] {$x_1x_2$} node [pos=0.7,below left,anchor=north east] {$x_5$} ++(135:1);
\matrix (mapa) [matrix of nodes,
        column sep={0.8cm,between origins},
        row sep={0.8cm,between origins},
        every node/.style={minimum size=0.3mm},
        anchor=4.center,
        ampersand replacement=\&] at (0.5,0.5)
{
                      \& |(c00)| 00         \& |(c01)| 01         \& |(c11)| 11         \& |(c10)| 10         \& |(cf)| \phantom{00} \\
|(r00)| 0             \& |(0)|  \phantom{0} \& |(1)|  \phantom{0} \& |(3)|  \phantom{0} \& |(2)|  \phantom{0} \&                     \\
|(r01)| 1             \& |(4)|  \phantom{0} \& |(5)|  \phantom{0} \& |(7)|  \phantom{0} \& |(6)|  \phantom{0} \&                     \\
|(rf) | \phantom{00}  \&                    \&                    \&                    \&                    \&                     \\
};
}%
{
\end{tikzpicture}
}

{
\begin{tikzpicture}[baseline=(current bounding box.north),scale=0.8]
\draw (0,0) grid (4,1);
\draw (0,1) -- node [pos=0.7,above right,anchor=south west] {$x_1x_2$} node [pos=0.7,below left,anchor=north east] {} ++(135:1);
\matrix (mapa) [matrix of nodes,
        column sep={0.8cm,between origins},
        row sep={0.7cm,between origins},
        every node/.style={minimum size=0.3mm},
        anchor=5.center,
        ampersand replacement=\&] at (0.5,0.8)
{
 \& |(c00)| 00         \& |(c01)| 01         \& |(c10)| 10         \& |(c11)| 11         \& |(cf)| \phantom{00} \\
 |(rf) | \phantom{00}   \& |(0)|  \phantom{0} \& |(1)|  \phantom{0} \& |(2)|  \phantom{0} \& |(3)|  \phantom{0} \&                     \\
};
}%
{
\end{tikzpicture}
}
{
\begin{tikzpicture}[baseline=(current bounding box.north),scale=0.8]
\draw (0,0) grid (4,1);
\draw (0,1) -- node [pos=0.7,above right,anchor=south west] {$x_1x_2$} node [pos=0.7,below left,anchor=north east] {} ++(135:1);
\matrix (mapa) [matrix of nodes,
        column sep={0.8cm,between origins},
        row sep={0.7cm,between origins},
        every node/.style={minimum size=0.3mm},
        anchor=5.center,
        ampersand replacement=\&] at (0.5,0.8)
{
 \& |(c00)| 00         \& |(c01)| 01         \& |(c11)| 11         \& |(c10)| 10         \& |(cf)| \phantom{00} \\
 |(rf) | \phantom{00}   \& |(0)|  \phantom{0} \& |(1)|  \phantom{0} \& |(3)|  \phantom{0} \& |(2)|  \phantom{0} \&                     \\
};
}%
{
\end{tikzpicture}
}

{
\begin{tikzpicture}[baseline=(current bounding box.north),scale=0.8]
\draw (0,0) grid (2,2);
\draw (0,2) -- node [pos=0.7,above right,anchor=south west] {$x_2$} node [pos=0.7,below left,anchor=north east] {$x_1$} ++(135:1);
\matrix (mapa) [matrix of nodes,
        column sep={0.8cm,between origins},
        row sep={0.8cm,between origins},
        every node/.style={minimum size=0.3mm},
        anchor=2.center,
        ampersand replacement=\&] at (0.5,0.5)
{
          \& |(c00)| 0          \& |(c01)| 1  \\
|(r00)| 0 \& |(0)|  \phantom{0} \& |(1)|  \phantom{0} \\
|(r01)| 1 \& |(2)|  \phantom{0} \& |(3)|  \phantom{0} \\
};
}%
{
\end{tikzpicture}
}

\usepackage{listings}
\usetikzlibrary{positioning, shapes, arrows.meta}

\usepackage[utf8x]{inputenc}

\definecolor{shellGreen}{RGB}{19,193,106}
\definecolor{backcolor}{rgb}{0.95,0.95,0.92}
\definecolor{mateBlack}{RGB}{45,45,50}
\definecolor{comment}{rgb}{0.1,0.6,0.2}
\definecolor{codegray}{rgb}{0.5,0.5,0.5}

\lstdefinestyle{verilog}{
   language=verilog,
   frame=single,
   basicstyle=\ttfamily\footnotesize,
   breaklines=true,
   captionpos=t,
   keepspaces=true,
   backgroundcolor=\color{backcolor},
   keywordstyle=[1]\color{blue}\bf,
   keywordstyle=[2]\color{red}\bf,
   keywordstyle=[3]\color{cyan!50}\bf,
   stringstyle=\color{orange},
   commentstyle=\color{comment},
   tabsize=2,
   showspaces=false,
   showstringspaces=false,
   showtabs=false,
   morekeywords=[1]{
      library, use ,all,entity,is,port,in,out,end,architecture,of, body,
      function, variable, begin,and,or,Not,downto,ALL, signal, process, if,
      else, elsif, case, when, then, range, to, component, type, with, select,
      others, constant, inout, buffer, map, true, false, array, subtype, wait,
      wait for, generic, =, <, >, <=, >=, =>, 
   },
   alsoletter={=, <, >},
   morekeywords=[2]{
          STD, textio, std_logic_textio, STD_LOGIC_VECTOR,STD_LOGIC,IEEE,STD_LOGIC_1164, work, local, real,
          math_real, time, NUMERIC_STD,STD_LOGIC_ARITH,STD_LOGIC_UNSIGNED,
          std_logic_vector, std_logic, ieee, numeric_std, std_ulogic,
          std_logic_1164, natural, bit, bit_vector, signed, unsigned,
          boolean, integer
    },
    morekeywords=[3]{rising_edge, falling_edge, resize, to_signed, to_unsigned},
    morecomment=[l]{--},
    rulecolor=\color{black},
}

\usepackage{diagbox}

\definecolor{codegreen}{rgb}{0,0.6,0}
\definecolor{codegray}{rgb}{0.5,0.5,0.5}
\definecolor{codepurple}{rgb}{0.58,0,0.82}
\definecolor{backcolour}{rgb}{0.95,0.95,0.92}

\lstdefinestyle{mystyle}{
    backgroundcolor=\color{backcolour}\bf,   
    commentstyle=\color{codegreen}\bf,
    keywordstyle=\color{magenta}\bf,
    numberstyle=\tiny\color{codegray}\bf,
    stringstyle=\color{codepurple},
    basicstyle=\ttfamily\footnotesize,
    breakatwhitespace=false,         
    breaklines=true,                 
    captionpos=t,                    
    keepspaces=true,                
    showspaces=false,                
    showstringspaces=false,
    showtabs=false,                  
    tabsize=2
}

\definecolor{vscode-bg}{HTML}{1E1E1E}      
\definecolor{vscode-text}{HTML}{D4D4D4}    
\definecolor{vscode-keyword}{HTML}{569CD6} 
\definecolor{vscode-string}{HTML}{D69D85}  
\definecolor{vscode-comment}{HTML}{6A9955} 
\definecolor{vscode-function}{HTML}{DCDCAA}
\definecolor{vscode-number}{HTML}{B5CEA8}  

\lstdefinestyle{vscode}{
    backgroundcolor=\color{vscode-bg},
    basicstyle=\ttfamily\footnotesize\color{vscode-text},
    keywordstyle=\color{vscode-keyword}\bfseries,
    commentstyle=\color{vscode-comment}\itshape,
    stringstyle=\color{vscode-string},
    numberstyle=\color{vscode-number},
    identifierstyle=\color{vscode-text},
    morekeywords={self}, 
    numbers=left,
    numberstyle=\tiny\color{gray},
    numbersep=5pt,
    frame=single,
    rulecolor=\color{gray},
    breaklines=true,
    showstringspaces=false,
    tabsize=4,
    captionpos=b,
    keepspaces=true,
    columns=fullflexible,
}

\definecolor{vscode-bg}{HTML}{FFFFFF}      
\definecolor{vscode-text}{HTML}{000000}    
\definecolor{vscode-keyword}{HTML}{0000FF} 
\definecolor{vscode-string}{HTML}{A31515}  
\definecolor{vscode-comment}{HTML}{008000} 
\definecolor{vscode-function}{HTML}{795E26}
\definecolor{vscode-number}{HTML}{098658}  

\lstdefinestyle{vscode-light}{
    backgroundcolor=\color{vscode-bg},
    basicstyle=\ttfamily\footnotesize\color{vscode-text},
    keywordstyle=\color{vscode-keyword}\bfseries,
    commentstyle=\color{vscode-comment}\itshape,
    stringstyle=\color{vscode-string},
    numberstyle=\color{vscode-number},
    identifierstyle=\color{vscode-text},
    morekeywords={self}, 
    numbers=left,
    numberstyle=\tiny\color{gray},
    numbersep=5pt,
    frame=single,
    rulecolor=\color{gray},
    breaklines=true,
    showstringspaces=false,
    tabsize=4,
    captionpos=b,
    keepspaces=true,
    columns=fullflexible,
    mathescape=true 
}

\definecolor{cnf-keyword}{RGB}{0,0,255}   
\definecolor{cnf-comment}{RGB}{0,128,0}   
\definecolor{cnf-number}{RGB}{255,0,0}    
\definecolor{cnf-operator}{RGB}{0,0,0}    

\lstdefinestyle{cnf}{
    backgroundcolor=\color{white},
    basicstyle=\ttfamily\footnotesize,
    keywordstyle=\color{cnf-keyword}\bfseries,
    commentstyle=\color{cnf-comment},
    numberstyle=\color{gray},
    numbersep=5pt,
    frame=single,
    breaklines=true,
    showstringspaces=false,
    tabsize=4,
    columns=fullflexible,
    morekeywords={p, cnf},          
    morecomment=[l]{c},             
    morestring=[b][\color{cnf-number}]0,  
    alsoletter={-},                 
    literate=%
    {-}{{\textcolor{cnf-operator}{-}}}1, 
    mathescape=true
}
    
\begin{document}
\bstctlcite{IEEEexample:BSTcontrol}

\title{Recurrent CircuitSAT Sampling for Sequential Circuits}

\makeatletter
    \newcommand{\linebreakand}{%
      \end{@IEEEauthorhalign}
      \hfill\mbox{}\par
      \mbox{}\hfill\begin{@IEEEauthorhalign}
    }
    \makeatother

\author{
    \IEEEauthorblockN{Arash Ardakani}
    \IEEEauthorblockA{
    \textit{University of California, Berkeley}\\
    arash.ardakani@berkeley.edu}
    \and
    \IEEEauthorblockN{Kevin He}
    \IEEEauthorblockA{
    \textit{University of California, Berkeley}\\
    kevinjhe@berkeley.edu}
    \and
    \IEEEauthorblockN{John Wawrzynek}
    \IEEEauthorblockA{
    \textit{University of California, Berkeley}\\
    johnw@berkeley.edu}
    }
    
\makeatletter
\patchcmd{\@maketitle}
  {\addvspace{0.5\baselineskip}\egroup}
  {\addvspace{-1\baselineskip}\egroup}
  {}
  {}
\makeatother


\newcommand\blfootnote[1]{%
  \begingroup
  \renewcommand\thefootnote{}\footnote{#1}%
  \addtocounter{footnote}{-1}%
  \endgroup
}

\definecolor{main}{HTML}{4472C4}    
\definecolor{sub}{HTML}{EBF4FF}     
\newcommand{\com}[1]{{\color{red}\sf{[#1]}}}
\newcommand{\OURS}{{\sc Demotic}}


\maketitle
\begin{abstract}
  In this work, we introduce a novel GPU-accelerated circuit satisfiability (CircuitSAT) sampling technique for sequential circuits. This work is motivated by the requirement in constrained random verification (CRV) to generate input stimuli to validate the functionality of digital hardware circuits. A major challenge in CRV is generating inputs for sequential circuits, along with the appropriate number of clock cycles required to meet design constraints. Traditional approaches often use Boolean satisfiability (SAT) samplers to generate inputs by unrolling state transitions over a fixed number of clock cycles. However, these methods do not guarantee that a solution exists for the given number of cycles. Consequently, producing input stimuli together with the required clock cycles is essential for thorough testing and verification. Our approach converts the logical constraints and temporal behavior of sequential circuits into a recurrent CircuitSAT problem, optimized via gradient descent to efficiently explore a diverse set of valid solutions, including their associated number of clock cycles. By operating directly on the circuit structure, our method reinterprets the sampling process as a supervised multi-output regression task. This differentiable framework enables independent element-wise operations on each tensor element, facilitating parallel execution during learning. As a result, we achieve GPU-accelerated sampling with substantial runtime improvements (up to $105.1\times$) over state-of-the-art heuristic samplers. We demonstrate the effectiveness of our method through extensive evaluations on circuit problems from the ISCAS-89 and ITC'99 benchmark suites.

\end{abstract}

\section{Introduction}

Simulation-based functional verification plays a crucial role in contemporary digital design automation processes, though it is often quite time-intensive \cite{Foster2015verification}. In this stage, the design undergoes thorough testing by running numerous simulations with a wide range of input signals to ensure that it complies with its intended functional requirements. For complex systems, each input typically spans numerous clock cycles, making exhaustive simulation impractical for real-world designs \cite{Bening2001verification}. Therefore, generating high-quality stimuli that effectively explore key functional scenarios, especially in corner cases, is essential for achieving sufficient coverage. Constrained random verification (CRV) \cite{Bhadra2007verification, Kitchen2007stimuli, Naveh2006crv, Yuan2010verification} addresses this challenge by allowing the user to define constraints that guide the generation of valid input stimuli, ensuring the design is tested in critical, bug-prone areas. CRV introduces randomness in the input selection process while still satisfying the given constraints, improving the efficiency of the verification process by increasing the likelihood of discovering hard-to-detect bugs \cite{Kitchen2007crv, Zhao2009crv, Naveh2013crv}.

The task of generating input stimuli in CRV for a given Boolean circuit is known as circuit satisfiability (CircuitSAT) sampling \cite{dutra2018quicksampler, Ardakani2025Demotic}. CircuitSAT sampling involves transforming circuit problems into Boolean satisfiability (SAT) problems and using SAT samplers to generate random solutions that meet the specified constraints. CircuitSAT is essentially a specialized form of the SAT problem \cite{Mishchenko2005Optimization, Tsai2009TimingAnalyzer, Bradley2011ModelChecking, Mishchenko2006EquivalenceChecking, Zhang2021LogicSynthesis}, where the circuit’s structure is represented explicitly as a network of logic gates rather than in conjunctive normal form (CNF). CircuitSAT sampling for combinational circuits focuses on generating diverse input patterns that satisfy the logic of the circuit, without involving any state or temporal dependencies. Since combinational circuits are acyclic and consist purely of logic gates (e.g., \textsc{and}, \textsc{or}, \textsc{not}), the sampling process is relatively straightforward and well-established in the literature \cite{dutra2018quicksampler, Ardakani2025Demotic, Ardakani2024diffsampler}. This process typically begins by converting the circuit’s structure into CNF using methods such as the Tseitin transformation \cite{tseitin1983complexity}. Any design constraints on the circuit's outputs can also be encoded in the CNF, after which SAT samplers can be applied to generate valid input stimuli for the CircuitSAT problem.

Unlike combinational circuits, sequential circuits depend on memory elements and clock-driven state transitions, making test generation significantly more complex. To apply SAT sampling to sequential circuits, the circuit must be ``unrolled'' over multiple time steps, effectively transforming it into a series of combinational steps that capture the circuit’s state transitions \cite{Chakraborty2020CircuitSAT}. However, this approach assumes that the required number of clock cycles for the given design constraints is known in advance, which is often unrealistic. Consequently, the CircuitSAT sampling process for sequential circuits involves not only generating input stimuli that satisfy the design constraints but also determining the necessary number of clock cycles for each input. This requirement makes CircuitSAT sampling for sequential circuits extremely challenging, yet invaluable if successfully implemented, as it provides a powerful means of testing and verification for complex, state-dependent systems.

In this work, we introduce a machine-learning-driven method to generate input stimuli for sequential circuits along with the necessary number of clock cycles for each sample. To this end, We formulate the CircuitSAT sampling process for sequential circuits as a recurrent multi-output regression task, where each logic gate is represented probabilistically and each memory element (e.g., flip-flops) is treated as hidden state, enabling the use of gradient descent (GD) to learn diverse solutions. This approach enables the parallel generation of independent input stimuli including their associated required clock cycles while accelerating the sampling computations with GPUs. We demonstrate the superior performance of our sampling method across all instances from the ISCAS-89 and ITC'99 benchmark suites. 

\section{Preliminaries}
\subsection{CircuitSAT Sampling} \label{subsec:circuitsat_sampling}
A digital circuit is an electronic system that processes binary information according to the rule of Boolean logic. Digital circuits are built using interconnected digital components including logic gates (e.g., \textsc{and}, \textsc{or}, and \textsc{not} gates) and memory elements (e.g., flip-flops). In digital circuits, variables are restricted to discrete binary values of either $0$ or $1$. The output of such circuits is produced based on how these digital components operate on binary variables. There are two types of digital circuits: combinational and sequential circuits. In combinational circuits, the output solely depends on current inputs, whereas in sequential circuits, the output depends not only on the current inputs but also on the history of inputs represented as the current state.

CircuitSAT sampling is the task of generating a sequence of a set of binary-valued assignments to the inputs of a given digital circuit that makes its output evaluate to desired value. The desired outputs are referred to as output constraints to the CircuitSAT problem. Sampling solutions from CircuitSAT instances in CRV is valuable as it enables efficient, targeted generation of test cases that satisfy specific logical constraints, ensuring that only valid and meaningful scenarios are explored. This approach increases coverage and detects edge cases more effectively than random sampling, focusing on diverse, relevant configurations that might reveal hidden bugs. CircuitSAT-based sampling also reduces the need for exhaustive testing, especially in complex circuits, by selectively exploring the solution space. Ultimately, this method improves verification efficiency and reliability, making it valuable for high-quality digital design testing.

A common method for CircuitSAT sampling is to use SAT solvers equipped with sampling features. These solvers not only check whether a Boolean formula is satisfiable but also sample specific solutions from the set of all possible solutions. Efficient SAT solving techniques include backtracking methods like the Davis-Putnam-Logemann-Loveland (DPLL) algorithm \cite{Davis1962DPLL}, stochastic local search approaches like WalkSAT \cite{selman1993local}, and conflict-driven clause learning (CDCL) algorithms \cite{Silva1996CDCL, silva2021CDCL}. In recent years, various algorithms for SAT and CircuitSAT sampling have been introduced, including randomized methods, Markov chain Monte Carlo (MCMC) approaches, and heuristic-based sampling techniques \cite{Impagliazzo2017RandomSAT, kitchen2009markov, Soos2020unigen3, dutra2018quicksampler, Golia2021cmsgen}. These techniques generally work by iteratively exploring possible solutions, selecting candidates based on certain rules, and using probabilistic criteria to accept or reject them.

The first step in CircuitSAT sampling using SAT solvers is to model the relationships among digital components and translate them into a CNF representation. CNF is structured as a conjunction (\textsc{and}) of multiple clauses, where each clause is a disjunction (\textsc{or}) of literals. Here, literals refer to Boolean variables or their negations. The number of variables in the CNF is determined by the primary inputs, intermediate signals, and primary outputs of the digital circuit. Conversely, the number of clauses corresponds to the logic operations in the circuit. The transformation into CNF offers SAT solvers a standardized representation of the problem, preserving the core constraints of the original circuit. The conversion to CNF is typically achieved using the Tseitin transformation \cite{tseitin1983complexity}.

Converting combinational circuits to CNF is straightforward, as each logic gate can be mapped to corresponding CNF sub-clauses using the Tseitin transformation. The size and intricacy of the resulting CNF formula can vary widely, influenced by factors like the number of gates, circuit depth, and counts of inputs, outputs, and intermediate signals. Generally, the CNF representation requires substantially more bit-wise operations than the original circuit. This added complexity increases the time SAT solvers need to find solutions due to the NP-complete nature of SAT problems. The challenge is even greater for sequential circuits, which involve time-dependent behavior. Representing these temporal dependencies in CNF necessitates additional variables and constraints to maintain consistency across time steps, further complicating the transformation. Although theoretically possible, converting sequential circuits to CNF involves unique difficulties such as state encoding, temporal logic, and clock handling, making it particularly challenging for large or complex designs. The main challenge in converting sequential circuits to CNF is determining the specific number of clock cycles to unroll the sequential circuit so that it can be treated as a combinational circuit for conversion using the Tseitin transformation. However, it is generally impossible to know the number of clock cycles that will guarantee satisfying solutions for given output constraints, creating a unique challenge for sequential CircuitSAT sampling.

\subsection{Multi-Output Regression Task}

A multi-output regression task involves predicting several target variables simultaneously based on a set of input features \cite{borchani2015survey}. The goal is to build a model that effectively captures the connections between the inputs and each output variable. This model can be developed through various approaches, such as linear regression or neural networks, and is trained on a dataset with known input-output pairs. During training, the model's parameters are optimized to reduce the difference between its predictions and the actual targets, often using mean squared error (MSE) or $\ell_2$ loss as a measure of accuracy.

\section{Methodology}
In contrast to combinational circuits, where outputs are determined solely by their present inputs, the output in sequential circuits depends on both the past behavior of the circuit and the present values of inputs. The temporal operations of sequential circuits are controlled by a clock signal. The contents of memory elements (i.e., flip-flops) represent the past behavior of such a circuit, which is commonly referred to as the \textit{state} of the circuit. 

\subsection{Recurrent CircuitSAT Sampler for Sequential Circuits} \label{sec:RCSampler}
Solving CircuitSAT problems for sequential circuits presents a unique challenge as it requires finding a sequence of inputs that satisfies the target output constraint over a series of clock cycles. To tackle such problems, we can leverage a novel technique inspired by recurrent neural networks (RNNs). In RNNs, backpropagation through time is utilized during the learning process, allowing for updates to the network's hidden state at each time step. Similarly, in the context of solving CircuitSAT problems, we perform forward computations to iteratively update the state values at each clock cycle. During backward computations, gradients are backpropagated through time, extending back to the initial time step (i.e., the first clock cycle), to adjust the input sequence accordingly. While this approach draws parallels to RNN training, it is tailored to the unique challenges posed by solving CircuitSAT problems for sequential circuits. 

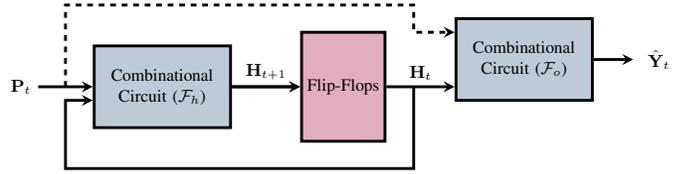
\begin{figure}[t]
    \centering
         \centering
         \scalebox{0.725}{\begin{tikzpicture}[auto, node distance=2cm,>=latex']
    \node [draw, fill = dateblue!30, shape=rectangle, minimum width=2.5cm, minimum height=1.5cm, text width=2cm, align = center, line width=1.5pt] (comb1) {Combinational Circuit ($\mathcal{F}_h$)};
    \node [draw, fill = datemagenta!40, shape=rectangle, right = 1.25cm of comb1, align = center, minimum height=2cm, line width=1.5pt] (ff) {Flip-Flops};
    \node [draw, fill = dateblue!30,shape=rectangle, minimum width=2.5cm, minimum height=1.5cm, right =1.25cm of ff, text width=2cm, align = center, yshift=0.5cm, line width=1.5pt] (comb2) {Combinational Circuit ($\mathcal{F}_o$)};
    \node [left =1cm of comb1, coordinate] (input1) {};
    \node [right of=comb2, coordinate] (output) {};
    \draw [->, >=stealth, line width=1.5pt] (input1) -- node[left, xshift = -0.5cm] {$\mathbf P_{t}$} (comb1);
    \draw [->, >=stealth, line width=1.5pt] (comb1) -- node {$\mathbf H_{t+1}$} (ff);
    \draw [->, >=stealth, line width=1.5pt] (comb1.east) -- node {} (ff);
    \draw [->, >=stealth, line width=1.5pt] (ff) -- node {$\mathbf H_{t}$} ([yshift=-0.5cm]comb2.west);
    \draw [->, >=stealth, line width=1.5pt] ([xshift=0.5cm]ff.east) |-  ([yshift=-1.5cm,xshift = -0.5cm]comb1.west) |-  ([yshift=-0.25cm, xshift = -0.15cm]comb1.west) |- ([yshift=-0.25cm]comb1.west);
    \draw [->, >=stealth, line width=1.5pt] (comb2) -- node[right, xshift = 0.5cm] {$\hat{\mathbf{Y}}_t$} (output);
    \draw [->, >=stealth, line width=1.5pt, dashed] ([yshift=0cm, xshift = -0.5cm]comb1.west) |- ([yshift=1cm, xshift = -0.75cm]comb2.west) |- ([yshift=0.5cm, xshift =0cm]comb2.west);
\end{tikzpicture}}
         \caption{The general form of a sequential circuit.}
         \label{fig2a}
         \vspace{-0.5cm}
\end{figure}


Fig. \ref{fig2a} shows the general structure of a sequential circuit. We use this structure to formulate the CircuitSAT problem for sequential circuits to find satisfying solutions. In this structure, there are two combination circuits: one to update the state of the circuit (i.e., the content values of flip-flops) and the other one to generate the output. It is worth mentioning that both of these combinational circuits take the present values of flip-flops and primary inputs at the current time step as their inputs. Given that these combinational circuits (i.e., $\mathcal{F}_h$ and $\mathcal{F}_o$) are fundamentally discrete and non-differentiable, we need to relax the CircuitSAT problem into a continuous version that still maintains the circuit's core structure and behavior. To do this, we utilize the probability model of digital gates, as outlined in Table \ref{tab1}. This probabilistic model is widely applied across various fields, including stochastic computing \cite{Ardakani2017SC} and the estimation of dynamic power in digital circuits \cite{harris2010cmos}. By applying these probabilities to model each gate, we produce a differentiable formulation of the circuit that preserves its 
functionality. Importantly, for any binary input, this model remains equivalent to the original circuit in its discrete form.

\begin{figure}[t]
    \centering
         \centering
         \scalebox{0.8}{\begin{tikzpicture}[scale=1, transform shape];
                \node [nnlayer]                     at ( 0,     0)    (sig1)          {$\mathcal{F}_o$};
                \node [nnlayer]                     at ( 1.5,   0)    (sig2)          {$\mathcal{F}_h$};

                \node [anchor=east]                 at ( -1,  -0.5) (hiddenlast)    {$\mathbf H_{t}$};
                \node [anchor=west]                 at ( 3.5,     1.) (hiddennext)    {$\mathbf H_{t+1}$};

                \node [anchor=west]                 at ( 0.6,     1.8) (hiddennext1)    {$\hat{\mathbf{Y}}_{t}$};

                \node [anchor=east]                 at ( -1, -1.5)  (input)         {$\mathbf P_{t}$};

                \draw [ultra thick, rounded corners=0.2cm] (hiddenlast) -| (sig1);
                \draw [ultra thick, rounded corners=0.2cm] (hiddenlast) -| (sig2);

                \draw [->, >=stealth, ultra thick, rounded corners=0.2cm] (sig1.north) -- ([yshift = 1.5cm]sig1.north);;


                \draw [->, >=stealth, ultra thick, rounded corners=0.2cm] (sig2.north) |- ([xshift=1.5cm, yshift=0.5cm]sig2.north);

                \draw [ultra thick, rounded corners=0.2cm] (input) -- ++(1.2,0) |- (0.5,-0.5);

                \begin{pgfonlayer}{background}
                \draw [fill=dateblue!10, rounded corners=.5cm] (-.5, -1) rectangle (3,1.5) node[anchor=south east, yshift=-2.5cm, xshift = -0.2cm] {Recurrent Cell};
                \end{pgfonlayer}

            \node [nnlayer,fill=dateblue!10]                     at ( -5,     0)    (RC)          {Recurrent Cell};
            
            \draw [<-, >=stealth, ultra thick, rounded corners=0.2cm] (RC.south) -- ([yshift = -1cm]RC.south) node[below] {$\mathbf P_{t}$};

            \draw [->, >=stealth, ultra thick, rounded corners=0.2cm] (RC.north) -- ([yshift = 1cm]RC.north) node[above] {$ \hat{\mathbf{Y}}_{t}$};

            \draw [->, >=stealth, ultra thick, dashed, rounded corners=0.2cm] (RC.east) -- ([xshift = 0.5cm]RC.east) -| ([xshift = 0.5cm, yshift= 0.5cm] RC.east) -- ([xshift = -0.5cm, yshift= 0.5cm] RC.east) -| ([xshift = -0.5cm] RC.west) -- (RC.west) node[below left] {$\mathbf H_{t+1}$};

            \end{tikzpicture}}
         \caption{Left: The proposed recurrent sequential model. Right: The proposed recurrent cell for sequential circuits.}
         \label{fig2b}
         \vspace{-0.5cm}
\end{figure}
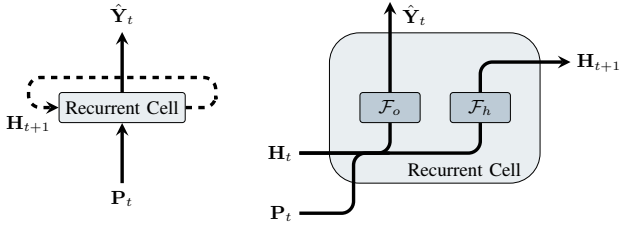

Let us represent the primary input variables at time step $t$ as $\textbf{V}_t \in \mathbb{R}^{b\times n}$. We encode the primary input variables at each time step as learnable parameters to an embedding layer followed by the sigmoid function to provide input probabilities at time $t$ as $\textbf{P}_t \in [0, 1]^{p\times n}$ to the combinational circuits, i.e.,
\begin{equation}
    \textbf{P}_t = \sigma(\textbf{V}_t).
\end{equation}
The present output of the circuit (i.e., $\hat{\textbf{Y}}_{t} \in [0, 1]^{b\times m}$) is computed as: 
\begin{equation}
    \hat{\textbf{Y}}_{t} = \mathcal{F}_o(\textbf{P}_t, \textbf{H}_{t}), \label{eq:output}
\end{equation}
where $\mathcal{F}_o$ and $\textbf{H}_t \in [0, 1]^{b \times r}$ denote the functionality of the combinational circuit generating outputs and the present values of flip-flops at each time step, respectively. The number of flip-flops in the circuit is represented by $r$. The state of the circuit for the next time step is obtained as:
\begin{equation}
    \textbf{H}_{t+1} = \mathcal{F}_h(\textbf{P}_t, \textbf{H}_{t}), \label{eq:state}
\end{equation}
where the functionality of the combinational circuit updating the values of flip-flops is denoted by $\mathcal{F}_h$. The $\ell_2$-loss function $\mathcal{L}$ can then be constructed by measuring the distance between $\hat{\textbf{Y}}_{t}$ at the desired time step $T$ and the target output valuation matrix $\textbf{Y} \in \{0, 1\}^{b \times m}$ as follows:
\begin{equation}
    \mathcal{L} = \sum_{b,m} \left|\left| \textbf{Y} - \hat{\textbf{Y}}_{T} \right|\right|^2_2.
\end{equation}
With such a formulation for sequential circuits, we can solve the CircuitSAT problem and provide $b$ solutions. The general form of the recurrent cell for sequential circuits is shown in Fig. \ref{fig2b}, which is analogous to the RNN cell.

With our formulation, backpropagation through time is used to compute the gradients of the loss with respect to the primary input variables (i.e., $\textbf{V}_t$). The objective is to find the gradient of $\mathcal{L}$ with respect to $\textbf{V}_t$. To derive this gradient, we apply the chain rule across time steps. The gradient with respect to $\textbf{V}_t$ requires accounting for how the circuit's states influence each other over time, computed as follows:
\begin{equation}
    \dfrac{\partial \mathcal{L}}{\partial \textbf{H}_t} = \dfrac{\partial \mathcal{L}}{\partial \hat{\textbf{Y}}_{t}} \cdot \dfrac{\partial \hat{\textbf{Y}}_{t}}{\partial \textbf{H}_t} + \dfrac{\partial \mathcal{L}}{\partial \textbf{H}_{t+1}} \cdot \dfrac{\partial \textbf{H}_{t+1}}{\partial \textbf{H}_t},
    \label{eq:BPTT}
\end{equation}
given Eq. (\ref{eq:output}) and Eq. (\ref{eq:state}). The recursive formula above captures the influence of the circuit state at each time step on both the output and the subsequent state. Backpropagation through time, as shown in Eq. (\ref{eq:BPTT}), can now be used to calculate the gradients of $\mathcal{L}$ with respect to $\textbf{V}_t$:
\begin{equation}
    \dfrac{\partial \mathcal{L}}{\partial \textbf{V}_t} = \dfrac{\partial \mathcal{L}}{\partial \hat{\textbf{Y}}_{t}} \cdot \dfrac{\partial \hat{\textbf{Y}}_{t}}{\partial \textbf{P}_t} \cdot \dfrac{\partial \textbf{P}_{t}}{\partial \textbf{V}_t} + \dfrac{\partial \mathcal{L}}{\partial \textbf{H}_{t+1}} \cdot \dfrac{\partial \textbf{H}_{t+1}}{\partial \textbf{P}_t} \cdot \dfrac{\partial \textbf{P}_{t}}{\partial \textbf{V}_t}.
\end{equation}
This formula enables updating the primary input variables by propagating the gradients back through time.

\begin{table}[]
    \centering
    \caption{Probability model and input derivatives of primary logic gates}
    \scalebox{0.85}{\begin{tabular}{c|c|c} 
\hline
\midrule
Operator  & Output Variable & Derivative w.r.t Input\\ 
 \midrule
\textsc{not}($P_{1}$) & $P_y = \overline{P_1} = 1 - P_{1}$ & $\dfrac{\partial P_y}{\partial P_1} = -1$  \\
\midrule
\textsc{and}($P_{1}$, $P_{2}$) & $P_y = P_{1}~P_{2}$ & $\dfrac{\partial P_y}{\partial P_1} = P_2$, $\dfrac{\partial P_y}{\partial P_2} = P_1$ \\
\midrule
\textsc{or}($P_{1}$, $P_{2}$) & $P_y = 1 - \overline{P_{1}}~\overline{P_{2}}$ & $\dfrac{\partial P_y}{\partial P_1} = \overline{P_{2}}$, $\dfrac{\partial P_y}{\partial P_2} = \overline{P_{1}}$\\
\midrule
\textsc{xor}($P_{1}$, $P_{2}$) & $P_y = \overline{P_{1}}~P_{2} + P_{1}~\overline{P_{2}}$ & $\dfrac{\partial P_y}{\partial P_1} = 1 -2P_2$, $\dfrac{\partial P_y}{\partial P_2} = 1 - 2 P_1$\\
\midrule
\textsc{xnor}($P_{1}$, $P_{2}$) & $P_y = P_{1}~P_{2} + \overline{P_{1}}~\overline{P_{2}}$ & $\dfrac{\partial P_y}{\partial P_1} = 2P_2-1$, $\dfrac{\partial P_y}{\partial P_2} = 2P_1-1$\\
\midrule
\hline

\end{tabular}}
    \label{tab1}
    \vspace{-0.5cm}
\end{table}

\subsection{Theoretical Analysis}
Proof of convergence is a crucial concept across various fields, including machine learning, economics, engineering, and operations research. Convergence in optimization is important because it ensures stable solutions, enhances robustness to noise, and enables the application of efficient algorithms that leverage gradient information and duality principles. In our formulation for the CircuitSAT problem, convergence plays a key role by simplifying the optimization process, thereby allowing our method to reliably find the satisfying solutions.

Before presenting a proof of convergence, let us first demonstrate that the optimization problem in our formulation is non-convex. The combinational parts of our formulation for the CircuitSAT problem can be viewed as an \textsc{and}-inverter graph (AIG), where the \textsc{and} operation is defined as
\begin{equation}
    \textsc{and}(p_1, p_2) = p_1 \cdot p_2,
\end{equation}
with \( p_1, p_2 \in [0, 1] \) as the inputs. To test the convexity of the \textsc{and} function, we analyze its Hessian matrix \( \mathcal{H} \) of second derivatives, which is given by
\begin{equation}
    \mathcal{H} = \begin{bmatrix}
    \dfrac{\partial^2 \textsc{and}(p_1, p_2)}{\partial p_1^2} & \dfrac{\partial^2 \textsc{and}(p_1, p_2)}{\partial p_1 \partial p_2} \\
    \dfrac{\partial^2 \textsc{and}(p_1, p_2)}{\partial p_2 \partial p_1} & \dfrac{\partial^2 \textsc{and}(p_1, p_2)}{\partial p_2^2}
    \end{bmatrix}
    = \begin{bmatrix}
    0 & 1 \\
    1 & 0
    \end{bmatrix}.
\end{equation}
The eigenvalues of \( \mathcal{H} \) are \( \pm 1 \). Since one eigenvalue is negative, \( \mathcal{H} \) is not positive semi-definite, and therefore, the \textsc{and} function is non-convex. As a result, the AIG containing multiple \textsc{and} gates is also non-convex.

While our CircuitSAT problem formulation is a non-convex optimization, we prove the convergence of gradient descent to global minima for such a non-convex formulation using the Lipschitz continuity of the gradient \cite{goldstein1977optimization}. Let us represent the sequence of inputs over \( T \) clock cycles as the matrix \( \textbf{X} \in [0, 1]^{n\times T} \) and let \( \mathcal{F}(\textbf{X}) \rightarrow [0, 1]^m \) be the non-convex function of the sequential circuit. Since \( \textbf{X} \) and \( \mathcal{F}(\textbf{X}) \) are bounded within \([0, 1]\), the $\ell_2$-loss ($\mathcal{L}(\textbf{X})$) of $\mathcal{F}(\textbf{X})$ is also bounded by the number of constrained outputs (i.e., $\mathcal{L}(\textbf{X}) \in [0, m]$) and, therefore, there exists a constant \( L > 0 \) such that 
\begin{equation}
    ||\nabla \mathcal{L}(\textbf{X}) - \nabla \mathcal{L}(\textbf{Z})|| \leq L||\textbf{X} - \textbf{Z}|| \quad \forall \textbf{X},\textbf{Z} \in [0, 1]^{n\times T},
\end{equation}
implying that the Hessian of the function \( \mathcal{L}(\textbf{X}) \) is bounded, i.e., \( \mathcal{H}(\textbf{X}) \leq L \). The gradient descent update for a matrix \( \textbf{X} \in [0, 1]^{n\times T} \) at the \( k \)th training iteration is as follows:
\begin{equation}
    \textbf{X}_{k+1} = \textbf{X}_{k} - \alpha \nabla\mathcal{L}(\textbf{X}_{k}),
    \label{eq:update}
\end{equation}
where \( \alpha > 0 \) denotes the learning rate. 

Using the second-order Taylor expansion of \( \mathcal{L}(\textbf{X}_{k+1}) \) around \( \textbf{X}_k \), we have:
\begin{align}
    \mathcal{L}(\textbf{X}_{k+1}) & \approx \mathcal{L}(\textbf{X}_{k}) + \nabla\mathcal{L}(\textbf{X}_{k})^T (\textbf{X}_{k+1} - \textbf{X}_{k}) \nonumber \\
    & + \dfrac{1}{2}(\textbf{X}_{k+1} - \textbf{X}_{k})^T H(\textbf{X}_{k})(\textbf{X}_{k+1} - \textbf{X}_{k}).
\end{align}
Using the Lipschitz bound on the Hessian, we can replace \( \mathcal{H}(\textbf{X}_k) \) with \( L \) to get the following inequality:
\begin{align}
    \mathcal{L}(\textbf{X}_{k+1}) & \leq \mathcal{L}(\textbf{X}_{k}) + \nabla\mathcal{L}(\textbf{X}_{k})^T (\textbf{X}_{k+1} - \textbf{X}_{k}) \nonumber \\
    & + \dfrac{L}{2}||\textbf{X}_{k+1} - \textbf{X}_{k}||^2.
\end{align}
Given Eq. (\ref{eq:update}), we can substitute \( \textbf{X}_{k+1} - \textbf{X}_{k} \) with \( -\alpha \nabla\mathcal{L}(\textbf{X}_{k}) \) and, therefore, rewrite the inequality as 
\begin{equation}
    \mathcal{L}(\textbf{X}_{k+1}) \leq \mathcal{L}(\textbf{X}_{k}) - \alpha ||\nabla\mathcal{L}(\textbf{X}_{k})||^2 + \dfrac{L\alpha^2}{2}||\nabla\mathcal{L}(\textbf{X}_{k})||^2.
\end{equation}
By defining \( \mu = \alpha - \dfrac{L\alpha^2}{2} \) for \( 0<\alpha<\dfrac{2}{L} \) and \( \mu >0 \), we have 
\begin{equation}
    \mathcal{L}(\textbf{X}_{k+1}) \leq \mathcal{L}(\textbf{X}_{k}) - \mu ||\nabla\mathcal{L}(\textbf{X}_{k})||^2.
\end{equation}
Since \( \mathcal{F}(\textbf{X}) \) is bounded, we can sum this inequality over \( K \in \{0, 1, \dots, K - 1\} \) and accordingly get:
\begin{equation}
    \mathcal{L}(\textbf{X}_{K}) \leq \mathcal{L}(\textbf{X}_{0}) - \mu \sum_{k=0}^{K-1}||\nabla\mathcal{L}(\textbf{X}_{k})||^2.
\end{equation}
By rearranging the above inequality, we have 
\begin{equation}
    \mu \sum_{k=0}^{K-1}||\nabla\mathcal{L}(\textbf{X}_{k})||^2 \leq \mathcal{L}(\textbf{X}_{0}) - \mathcal{L}(\textbf{X}_{K}) \leq \mathcal{L}(\textbf{X}_{0}).
\end{equation}
Since \( \mathcal{L}(\textbf{X}_{0}) \) is bounded within \([0, 1]\), we can conclude
\begin{equation}
    \sum_{k=0}^{K-1}||\nabla\mathcal{L}(\textbf{X}_{k})||^2 < \infty.
\end{equation}
The above inequality implies that \( ||\nabla\mathcal{L}(\textbf{X}_{k})||^2 \rightarrow 0 \) as \( k \rightarrow \infty \), showing the gradient descent algorithm converges to a stationary point where \( \nabla\mathcal{L}(\textbf{X}) = 0 \). Since \( \mathcal{L}(\textbf{X}) \) is bounded and non-convex, gradient descent guarantees convergence to a stationary point within \([0, 1]\) for inputs, though this may be a global minimum, local minimum or saddle point. 


Given our probabilistic formulation, $\nabla\mathcal{L}(\textbf{X})$ can only be zero in two cases. The first occurs when the $\ell_2$-loss (i.e., $\mathcal{L}(\textbf{X})$) is zero. This implies that the output constraints are satisfied for the given inputs, and the inputs to the circuit must have converged to either $0$ or $1$, as the outputs can only take discrete binary values when the inputs are also discrete binary values in our formulation. This guarantees that the inputs associated with a loss value of zero are satisfying solutions and, as such, represent global minima. 

The second case in which $\nabla\mathcal{L}(\textbf{X}) = 0$ occurs when the logic gates cause the gradients to be zero through their derivative functions with respect to their inputs (see Table \ref{tab1}). For example, a single \textsc{and} gate cannot reach an output of $1$ through gradient descent if both of its inputs are initialized to zero. This results in a non-zero loss while $\nabla\mathcal{L}(\textbf{X})$ remains zero, which is referred to as a saddle point. However, this scenario is unlikely (though possible) to occur since the inputs are randomly initialized within $[0, 1]$. Moreover, even if it does happen, it can be mitigated by regularization techniques such as input decay and noise injection.

\subsection{Backtracking Algorithm}
Given our differentiable recurrent sampling method described in Section \ref{sec:RCSampler}, we now introduce a backtracking algorithm that can generate satisfying input solutions of varying lengths based on specific design constraints. All generated solutions are constrained to have their initial state values set to zero. We begin the sampling process by defining the upper bound for the number of clock cycles, denoted as $\eta$. The goal is to generate solutions where the required number of clock cycles for each sample is less than or equal to this upper bound. Starting with the design constraint on the output of a sequential circuit, we run the differentiable recurrent sampling method for one clock cycle and generate potential solutions. If the generated solutions meet the design constraint, they are stored; otherwise, they are discarded. This process is repeated for $2$ clock cycles, continuing the same procedure. The cycle-by-cycle sampling process continues incrementally until reaching the specified upper limit on clock cycles. By then, we will have solutions that require varying numbers of clock cycles. It is possible, however, that for certain numbers of clock cycles, no solutions are found, as the constraints may not be satisfiable for those specific cycle lengths. Without loss of generality, a lower bound can also be defined for the number of clock cycles, ensuring that the generated solutions fall within a user-specified range. Algorithm \ref{alg1} summarizes the backtracking algorithm.

\begin{algorithm}[t]
\caption{Pseudo Code of our Backtracking Algorithm}
\begin{algorithmic}[1]
    \STATE \textbf{Input:} The upper bound for clock cycles $\eta$, the number of training iterations $itr$, the target output valuation matrix $\textbf{Y}$ 
    \STATE \textbf{Output:} List of satisfying primary inputs $PI$
    \FOR{$cc = 1$ to $\eta$}
            \FOR{$j = 1$ to $itr$}
                \FOR{$t = 1$ to $cc$}
                    \STATE $\textbf{P}_t = \sigma(\textbf{V}_t)$
                    \STATE $\textbf{Y}_{t} = \mathcal{F}_o(\textbf{P}_t, \textbf{H}_{t-1})$
                    \STATE $\textbf{H}_{t} = \mathcal{F}_h(\textbf{P}_t, \textbf{H}_{t-1})$
                \ENDFOR
                \STATE $\mathcal{L} = \sum_{b,m} \left|\left| \textbf{Y} - \hat{\textbf{Y}}_{cc} \right|\right|^2_2$
                \STATE $\mathcal{L}$.backward()
                \STATE optimizer.step()
                \IF{$\textbf{P}$ is satisfiable}
                    \STATE Append \textbf{P} to $PI$ 
                \ENDIF    
            \ENDFOR
    \ENDFOR
    \STATE \textbf{Return} $PI$
    \vspace{-0.65cm}
\end{algorithmic}

\label{alg1}
\end{algorithm}

\section{Experimental Results}
In this section, we showcase the effectiveness of our sampling approach. To this end, we implemented a prototype of our method using PyTorch, an open-source library that merges Torch's GPU-optimized backend with a Python-compatible interface. For a comprehensive evaluation, we utilize the complete ISCAS-89 and ITC'99 benchmark suites, incorporating all components. Our experiments were conducted on a system with an Intel Xeon E$5$-$2698$ processor running at $2.2$ GHz and $8$ NVIDIA V$100$ GPUs, each equipped with $32$ GB of memory. We assess the runtime performance of our method based on throughput, defined as the number of unique and valid solutions produced per second. GD was used as the optimizer during training, with a learning rate of $50$ and a total of $5$ iterations. Depending on the size of each CircuitSAT instance, we adjust the batch size between $100$ and $1,000,000$. It is also noteworthy that the maximum sample count generated by our sampler is limited by the product of the batch size and the maximum number of clock cycles (i.e., $\eta$). For our experiments, we set the maximum clock cycle limit to $\eta = 50$.

\subsection{Runtime Performance}
In sequential circuit sampling, the goal is to identify input sequences of varying lengths (i.e., requiring different numbers of clock cycles) that satisfy given output constraints. To transform these circuits into CircuitSAT sampling problems, we randomly assign specific binary values to certain output nodes. Table \ref{tab3} presents the sampling performance of our recurrent CircuitSAT sampler in terms of throughput, defined as the number of unique input sequences generated per second, for a representative set of $16$ sequential CircuitSAT instances from the ISCAS-89 and ITC'99 benchmark suites. The generated samples are input sequences with clock cycles ranging from $1$ to $50$. Producing valid input sequences of different lengths is essential for sequential CircuitSAT sampling, as specific output constraints may preclude solutions at certain clock cycle lengths. Furthermore, samples with varying sequence lengths (i.e., clock cycles) enable the detection of both static logic faults and dynamic issues related to timing and state transitions.

\begin{table*}[t]
    \centering
    \caption{The runtime performance of sampler is evaluated in terms of unique solution throughput for sequential circuits.}
    \scalebox{0.92}{\begin{tabular}{c|c|c|c|c|c|c|c|c} 
\hline
\midrule
CircuitSAT  & \# Primary  & \# Primary & \# Logic  & \multirow{ 2}{*}{\# DFF} & \# Variables  & \# Clauses & \multicolumn{2}{c}{Throughput (\# Unique Solutions per Second)} \\ 
 Instance & Inputs & Outputs & Gates &  & (CNF) & (CNF) & \textbf{This work} & {\sc CMSGen} \\ 
 \midrule
s$208.1$    & $10$   & $1$   & $104$    & $8$   & $7696$   & $7371$  & $\textbf{114,242.4}~(105.1\times)$  & $1087.2$ \\
s$386$   & $7$  & $7$   & $159$  & $6$  & $12,790$  & $12,609$ & $\textbf{35,849.5}~(65.6\times)$ & $546.6$ \\
s$526$   & $3$  & $6$  & $193$  & $21$  & $16,772$  & $16,692$ & $\textbf{15,978.6}~(32.6\times)$ & $490.8$ \\
s$832$   & $18$  & $19$  & $287$  & $5$ & $25,386$ & $24,919$   & $\textbf{5,700.4}~(30.5\times)$ & $186.9$ \\
s$1196$  & $14$  & $14$  & $529$  & $18$ & $38,592$ & $38,230$   & $\textbf{5,651.1}~(51.7\times)$ & $109.4$ \\
s$4863$  & $49$  & $16$  & $2342$  & $104$ & $166,035$ & $164,796$   & $\textbf{157.6}~(75.0\times)$ & $2.1$ \\
s$15850.1$  & $77$ & $150$ & $9772$ & $534$ & $606,514$ & $604,444$   & $\textbf{14.5}~(3.7\times)$ & $3.9$ \\
s$38584$  & $12$  & $278$  & $19,253$ & $1,452$ & $1,357,767$ & $1,357,194$    & $\textbf{1.7}~(2.1\times)$ & 0.8 \\

\midrule

b$01$ & $2$ & $2$ & $45$ & $5$ & $3,205$ & $3,154$ & $\textbf{54,774.3}~(18.2\times)$ & $3,012.6$ \\
b$06$ & $2$ & $6$ & $56$ & $9$ & $3,271$ & $3,216$ & $\textbf{173,909.9}~(66.5\times)$ & $2,614.3$ \\
b$10$ & $11$ & $6$ & $206$ & $17$ & $13,963$ & $13,683$ & $\textbf{5,463.6}~(13.8\times)$ & $396.8$ \\
b$12$ & $5$ & $6$ & $1,076$ & $121$ & $78,559$ & $78,429$ & $\textbf{402.5}~(4.1\times)$ & $98.2$ \\
b$15$ & $36$ & $70$ & $8,922$ & $449$ & $660,026$ & $659,059$ & $\textbf{3.1}~(15.5\times)$ & $0.2$ \\
b$17$ & $37$ & $97$ & $32,326$ & $1,415$ & $2,379,165$ & $2,378,145$ & $\textbf{0.2}$ & TO \\
b$19$ & $24$ & $30$ & $231,320$ & $6,642$ & $16,819,969$ & $16,819,339$ & $\textbf{0.01}$ & TO  \\
b$22$ & $32$ & $22$ & $29,951$ & $735$ & $2,183,304$ & $2,182,485$ & $\textbf{0.5}~(5\times)$ & $0.1$ \\
\midrule
\hline
\end{tabular}}
    \label{tab3}
    \vspace{-0.5cm}
\end{table*}

\begin{figure}[t]
    \centering
    \input{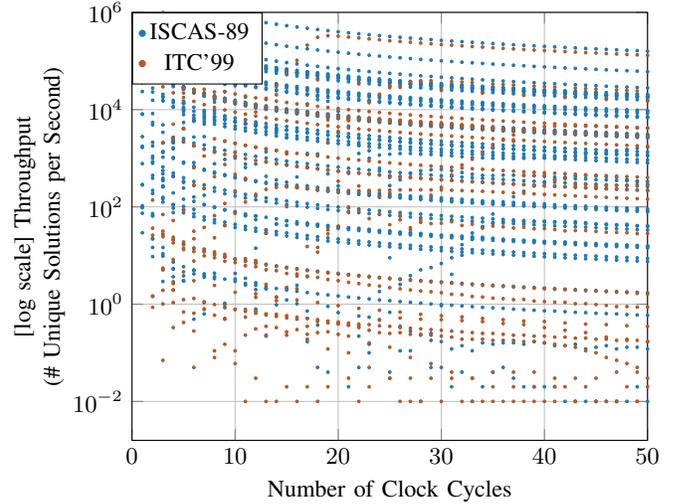}
    \caption{A log plot showing the runtime  performance of our recurrent sampler in terms of throughput, measured as the number of unique satisfying solutions per second, versus the number of clock cycles across all instances from the ISCAS-89 and ITC'99 benchmark suites. The throughput values are rounded to two decimal places in this plot.}
    \label{fig2}
    \vspace{-0.5cm}
\end{figure}

To compare runtime performance with previous works, we unrolled the sequential circuits in the ISCAS-89 and ITC'99 benchmark suites over $25$ clock cycles, converting the resulting combinational circuit to CNF using the Tseitin transformation. For this CircuitSAT sampling process, we applied the same output constraints used in our recurrent sampler experiments. These output constraints were initially set to ensure the existence of solutions at $25$ clock cycles. We selected $25$ clock cycles as it represents the average of all clock cycles (ranging from $1$ to $50$) used in our sampling experiments. 

We compare the runtime performance of our recurrent CircuitSAT sampler with state-of-the-art SAT sampling methods, specifically {\sc UniGen3} \cite{Soos2020unigen3}, {\sc CMSGen} \cite{Golia2021cmsgen}, and {\sc DiffSampler} \cite{Ardakani2024diffsampler}, each tasked with generating at least $1,000$ solutions within a $4$-hour timeframe. While these samplers work directly on the CNF representations of CircuitSAT instances, our sampler operates on the sequential circuits themselves, preserving their original structure and temporal behavior. Both {\sc UniGen3} and {\sc CMSGen} are optimized C++ implementations, while {\sc DiffSampler} is a Python-based SAT sampler that leverages JAX for GPU acceleration. {\sc UniGen3} and {\sc CMSGen} were tested on a server-grade AMD EPYC $9274$F CPU with the maximum frequency of $4.3$ GHz and $1$ TB of RAM, whereas {\sc DiffSampler} was evaluated on a system with an Intel Xeon E$5$-$2698$ processor at $2.2$ GHz and $8$ NVIDIA V$100$ GPUs, each with $32$ GB of memory.

Our experiments show that {\sc UniGen3} and {\sc DiffSampler} can only generate solutions for the CircuitSAT instance ``s$27$'' with throughputs of $58,180.0$ and $64,458.4$, respectively. For this specific CircuitSAT instance, our recurrent sampler outperforms both {\sc UniGen3} and {\sc DiffSampler} by factors of $4.3\times$ and $3.9\times$, respectively. On the other hand, {\sc CMSGen} generates solutions across the majority of CircuitSAT instances, as shown in Table~\ref{tab3}. According to Table~\ref{tab3}, our recurrent CircuitSAT sampler achieves up to a $105.1\times$ increase in throughput compared to {\sc CMSGen}. Furthermore, our recurrent sampler consistently outperforms {\sc CMSGen}, achieving an average speedup of $24.3\times$ across all instances for which {\sc CMSGen} generated solutions. Figure \ref{fig2} illustrates the throughput of our sampler across various clock cycle counts for all instances from the ISCAS-89 and ITC'99 benchmark suites, with throughput values rounded to two decimal places in the plot. This figure demonstrates that our sampler can produce input sequences of different lengths for a given CircuitSAT problem. Notably, some instances lack solutions that meet the specified output constraints at certain clock cycles.

For conventional samplers, including {\sc CMSGen}, {\sc UniGen3}, and {\sc DiffSampler}, we fixed the number of clock cycles at $25$ to ensure a fair runtime performance comparison. However, these samplers can also be used to generate input sequences over varying clock cycles. This approach, however, requires transforming the sequential circuit into a separate CNF for each clock cycle, which leads to an increase in CNF size with the number of clock cycles. In contrast, our method recursively computes forward computations for any desired number of clock cycles (see Algorithm \ref{alg1}) and calculates gradients with respect to inputs as described in Section \ref{sec:RCSampler}.

\subsection{Learning Dynamics}
Here, we present the learning dynamics of our recurrent sampler for a few representative sequential CircuitSAT instances. We begin by analyzing the distribution of the generated input sequences across $50$ clock cycles. The randomly specified output constraints ensure the existence of satisfying solutions at the $25$th clock cycle for runtime comparisons with prior works. However, there is no guarantee of solutions existing for other clock cycles. For example, Fig. \ref{fig3} highlights three representative sequential CircuitSAT instances, showcasing the percentage of valid and unique solutions generated by our sampler (calculated as the number of unique solutions divided by the batch size) for specific clock cycles.

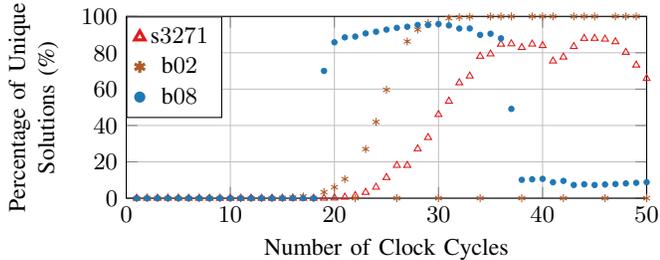
\begin{figure}[t]
    \vspace{-0.25cm}
    \centering
    \begin{tikzpicture}
    \begin{axis}[
        xlabel={Number of Clock Cycles},
        ylabel={Percentage of Unique Solutions (\%)},
        ylabel style={align=center, text width=3cm},
        xmin=0, xmax=50,
        ymin=0, ymax=100,
        grid=both,
        grid style={line width=.1pt, draw=gray!10},
        major grid style={line width=.2pt,draw=gray!50},
        xlabel near ticks,
        tick align=inside,
        legend columns=3,
        transpose legend,
        height=4cm, width=8.5cm,
        legend style={mark options={scale=1.5, very thick}, at={(0,1)}, anchor=north west},
    ]
    \addplot[only marks, mark=triangle, color=Paired-5, mark size=1.5pt] coordinates {

(1,  0/150)
(2,  2/150)
(3,  6/150)
(4,  10/150)
(5,  13/150)
(6,  2/150)
(7,  3/150)
(8,  0/150)
(9,  0/150)
(10, 0/150)
(11, 0/150)
(12, 0/150)
(13, 0/150)
(14, 0/150)
(15, 0/150)
(16, 0/150)
(17, 0/150)
(18, 0/150)
(19, 5/150)
(20, 19/150)
(21, 87/150)
(22, 221/150)
(23, 484/150)
(24, 911/150)
(25, 1706/150)
(26, 2710/150)
(27, 2710/150)
(28, 4078/150)
(29, 5006/150)
(30, 6908/150)
(31, 8007/150)
(32, 9524/150)
(33, 10084/150)
(34, 11711/150)
(35, 11906/150)
(36, 12710/150)
(37, 12740/150)
(38, 12434/150)
(39, 12718/150)
(40, 12593/150)
(41, 11319/150)
(42, 11647/150)
(43, 12517/150)
(44, 13188/150)
(45, 13194/150)
(46, 13128/150)
(47, 12911/150)
(48, 12038/150)
(49, 10985/150)
(50, 9884/150)

    };
    \addlegendentry{s$3271$}

    \addplot[only marks, mark=asterisk, color=Paired-11, mark size=1.5pt] coordinates {

(1, 0 )
(2, 0 )
(3, 8/1000 )
(4, 8/1000 )
(5, 8/1000 )
(6, 16/1000 )
(7, 8/1000 )
(8, 8/1000 )
(9, 8/1000 )
(10,0 )
(11,8/1000 )
(12,8/1000 )
(13,32/1000 )
(14,0 )
(15,128/1000 )
(16,339/1000 )
(17,954/1000 )
(18,0 )
(19,3315/1000 )
(20,6042/1000 )
(21,10506/1000 )
(22,0 )
(23,27030/1000 )
(24,41983/1000 )
(25,59645/1000 )
(26,0 )
(27,86219/1000 )
(28,92755/1000 )
(29,96284/1000 )
(30,0 )
(31,99046/1000 )
(32,99535/1000 )
(33,99740/1000 )
(34,0 )
(35,99944/1000 )
(36,99968/1000 )
(37,99983/1000 )
(38,0 )
(39,99996/1000 )
(40,99997/1000 )
(41,99998/1000 )
(42,0 )
(43,100000/1000 )
(44,100000/1000 )
(45,100000/1000 )
(46,0 )
(47,100000/1000 )
(48,100000/1000 )
(49,100000/1000 )
(50,0 )

    };
    \addlegendentry{b$02$}

    \addplot[only marks, mark=*, color=Paired-1, mark size=1.pt,] coordinates {
(1,  0/1000)
(2,  0/1000)
(3,  0/1000)
(4,  0/1000)
(5,  0/1000)
(6,  0/1000)
(7,  0/1000)
(8,  0/1000)
(9,  0/1000)
(10, 0/1000)
(11, 0/1000)
(12, 0/1000)
(13, 0/1000)
(14, 0/1000)
(15, 0/1000)
(16, 0/1000)
(17, 0/1000)
(18, 0/1000)
(19, 70013/1000)
(20, 85762/1000)
(21, 88479/1000)
(22, 88928/1000)
(23, 90654/1000)
(24, 91586/1000)
(25, 92735/1000)
(26, 93725/1000)
(27, 94311/1000)
(28, 95315/1000)
(29, 95286/1000)
(30, 95834/1000)
(31, 95052/1000)
(32, 93342/1000)
(33, 93375/1000)
(34, 89838/1000)
(35, 90487/1000)
(36, 87927/1000)
(37, 49134/1000)
(38, 10117/1000)
(39, 10382/1000)
(40, 10643/1000)
(41, 8753/1000)
(42, 9527/1000)
(43, 7276/1000)
(44, 7626/1000)
(45, 7247/1000)
(46, 7539/1000)
(47, 7785/1000)
(48, 8115/1000)
(49, 8460/1000)
(50, 8791/1000)
    };
    \addlegendentry{b$08$}

    \addplot[only marks, mark=o, color=Paired-1, mark size=1.5pt] coordinates {

    };
    \addlegendentry{This work}


    
    \end{axis}
\end{tikzpicture}
    \caption{A plot illustrating the percentage of unique solutions generated by our recurrent sampler over $50$ clock cycles for three representative CircuitSAT instances. The batch size for this experiment is set to $100,000$.
}
    \label{fig3}
    \vspace{-0.5cm}
\end{figure}

Fig. \ref{fig4} shows the learning progress of our recurrent sampler, showing the total number of unique solutions generated over five iterations. The learning curves indicate that as the number of iterations increases, the total number of unique solutions identified by our sampler over $50$ clock cycles also grows. This aligns with our theoretical convergence proof, which shows that gradient descent can reach a global minimum, achieving zero-valued loss in the non-convex landscapes of our continuous formulation for sequential CircuitSAT problems. Of course, the convergence rate varies across CircuitSAT instances, influenced by the complexity and structure of the sequential circuit, as well as the hyper-parameters selected for the learning process.

\section{Related Work}

Over the years, researchers have developed various CircuitSAT/SAT sampling strategies. Among these, tools like {\sc UniGen3} aim to maintain approximate uniformity in sampling \cite{yash2022barbarik}, while {\sc CMSGen} and {\sc Quicksampler} \cite{dutra2018quicksampler} prioritize speed and efficiency. Efforts to exploit data-parallel hardware for SAT solving have largely revolved around accelerating CDCL and heuristic-based methods \cite{costa2013parallelization, osama2021sat}. More recently, approaches such as {\sc MatSat} \cite{sato2021matsat} and {\sc NeuroSAT} \cite{amizadeh2018learning} have reimagined SAT problems as constrained optimization tasks over continuous domains. However, these innovations have primarily targeted small, random benchmarks, leaving their potential for large-scale and standard benchmarks unrealized, particularly with GPU-accelerated sampling. The introduction of {\sc DiffSampler} \cite{Ardakani2024diffsampler} marks a turning point by demonstrating competitive performance in GPU-accelerated SAT sampling on widely used benchmarks, rivaling methods like {\sc UniGen3} and {\sc CMSGen}. Complementing this, {\sc Demotic} \cite{Ardakani2025Demotic} offers a GPU-powered, sampling solution tailored for combinational CircuitSAT in CRV, operating directly on hardware description language representations such as Verilog.

Despite the high runtime performance of the aforementioned CircuitSAT samplers, they are primarily designed for combinational circuits. Recently, a symbolic algorithm based on Algebraic Decision Diagrams, called {\sc TraceSampler}, has been introduced to sample bounded traces (i.e., sequences of states) in sequential circuits with guaranteed uniformity \cite{Chakraborty2020CircuitSAT}. However, {\sc TraceSampler} focuses on uniform sampling of fixed-length traces, typically starting from a defined initial state, without applying constraints on the circuit’s primary outputs.

Unlike the previous sampling approaches, our sampler takes a fundamentally different approach. Our formulation reframes the sampling procedure as an optimization task that can be minimized using gradient descent under specified output constraints, with a theoretical proof of convergence to a stationary point. During optimization, forward and backward computations are performed recurrently over any number of clock cycles without unrolling the circuit (i.e., converting it to a combinational circuit). This allows sampling from the solution space across varying clock cycles, which is crucial for detecting timing-related and state-transition faults. Additionally, due to the parallel nature of this optimization process across different data batches, our method can be accelerated using GPUs. Consequently, our sampler outperforms the state-of-the-art (i.e., {\sc CMSGen}) in terms of runtime performance across all CircuitSAT instances from the ISCAS-89 and ITC'99 benchmark suites.

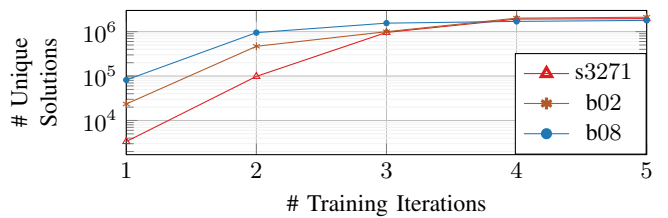
\begin{figure}[t]
    \centering
    \begin{tikzpicture}
    \begin{axis}[
        xlabel={\# Training Iterations},
        ylabel={\# Unique Solutions},
        ylabel style={align=center, text width=2cm},
        xmin=1, xmax=5,
        ymin=0, ymax=3000000,
        ymode=log,
        grid=both,
        grid style={line width=.1pt, draw=gray!10},
        major grid style={line width=.2pt,draw=gray!50},
        xlabel near ticks,
        tick align=inside,
        legend columns=3,
        transpose legend,
        height=3.5cm, width=8.5cm,
        legend style={mark options={scale=1.5, very thick}, at={(1,0)}, anchor=south east},
    ]
    \addplot[mark=triangle, color=Paired-5, mark size=1.5pt] coordinates {

(1, 3357)
(2, 98315)
(3, 960484)
(4, 1918341)
(5, 1965772)

    };
    \addlegendentry{s$3271$}

    \addplot[mark=asterisk, color=Paired-11, mark size=1.5pt] coordinates {

(1,  23519)
(2,  468405)
(3,  997364)
(4,  2015149)
(5,  2111032)

    };
    \addlegendentry{b$02$}

    \addplot[ mark=*, color=Paired-1, mark size=1.pt,] coordinates {
(1, 81771 )
(2, 950911 )
(3, 1554177 )
(4, 1700901 )
(5, 1794044 )
    };
    \addlegendentry{b$08$}

    \addplot[only marks, mark=o, color=Paired-1, mark size=1.5pt] coordinates {

    };
    \addlegendentry{This work}


    
    \end{axis}
\end{tikzpicture}
    \caption{A log plot illustrating the number of unique solutions generated by our recurrent sampler over $5$ training iterations for three representative CircuitSAT instances. 
}
    \label{fig4}
    \vspace{-0.5cm}
\end{figure}
\section{Conclusion}
In this work, we introduced a novel GPU-accelerated CircuitSAT sampling approach tailored for sequential circuits, addressing key challenges in CRV. By reframing the sampling process as a recurrent optimization problem and utilizing gradient descent, our method efficiently generates input stimuli along with the required clock cycles, ensuring compliance with design constraints. We provided theoretical guarantees showing that minimizing this recurrent optimization process via gradient descent converges to a stationary point, with a high likelihood of reaching a global minimum by incorporating regularization techniques to avoid saddle points. Operating directly on the circuit structure without unrolling state transitions, our approach redefines sampling as a differentiable multi-output regression task, enabling highly parallel computations and delivering substantial runtime improvements. Extensive evaluations on benchmark suites such as ISCAS-89 and ITC'99 demonstrate the effectiveness and scalability of our method, achieving up to $105.1\times$ higher throughput compared to state-of-the-art samplers. This work lays the foundation for more efficient verification methods for sequential systems and paves the way for integrating differentiable sampling techniques into broader hardware design and validation processes.

\bibliographystyle{IEEEtran}
\bibliography{sat_sampling.bib}

\end{document}